\documentstyle[11pt,adassconf]{article}  % Leave intact

\begin{document}   % Leave intact

%-----------------------------------------------------------------------
%			    Paper ID Code
%-----------------------------------------------------------------------
% Enter the proper paper identification code.  The ID code for your
% paper is the session number associated with your presentation as
% published in the official ADASS 2000 conference program.  You can
% find this number locating your abstract in the printed program
% that you received at the meeting or on-line via
% http://hea-www.harvard.edu/ADASS/; the ID code is the letter-number
% sequence proceeding the title of your presentation.
%
% This will not appear in your paper; however, it allows different
% papers in the proceedings to cross-reference each other.
%
% EXAMPLE: \paperID{O1-03}
% EXAMPLE: \paperID{P2-56}
%
% Note that you should only have one \paperID, and it should not
% include a trailing period.  

\paperID{P2-31}

%-----------------------------------------------------------------------
%		            Paper Title 
%-----------------------------------------------------------------------
% Enter the title of the paper.
%
% EXAMPLE: \title{A Breakthrough in Astronomical Software Development}
% 

\title{DISCoS - a detector-independent software for the on-ground testing and calibration of scientific payloads using the ESA Packet Telemetry and Telecommand Standards }

%-----------------------------------------------------------------------
%		          Authors of Paper
%-----------------------------------------------------------------------
% Enter the authors followed by their affiliations.  The \author and
% \affil commands may appear multiple times as necessary (see example
% below).  List each author by giving the first name or initials first
% followed by the last name.  Authors with the same affiliations
% should grouped together. 
%
% EXAMPLE: \author{Raymond Plante, Doug Roberts, 
%                  R.\ M.\ Crutcher\altaffilmark{1}}
%          \affil{National Center for Supercomputing Applications, 
%                 University of Illinois Urbana-Champaign, Urbana, IL
%                 61801}
%          \author{Tom Troland}
%          \affil{University of Kentucky}
%
%          \altaffiltext{1}{Astronomy Department, UIUC}
%
% In this example, the first three authors, "Plante", "Roberts", and
% "Crutcher" are affiliated with "NCSA".  "Crutcher" has an alternate 
% affiliation with the "Astronomy Department".  The fourth author,
% "Troland", is affiliated with "University of Kentucky"

\author{F. Gianotti and M. Trifoglio }
\affil{Istituto TeSRE/CNR -Via P. Gobetti 101, 40129 Bologna, Italy }

%-----------------------------------------------------------------------
%			 Contact Information
%-----------------------------------------------------------------------
% This information will not appear in the paper but will be used by
% the editors in case you need to be contacted concerning your
% submission.  Enter your name as the contact along with your email
% address.
% 
% EXAMPLE:  \contact{Dennis Crabtree}
%           \email{crabtree@cfht.hawaii.edu}
%

\contact{Massimo Trifoglio }
\email{trifoglio@tesre.bo.cnr.it}

%-----------------------------------------------------------------------
%		      Author Index Specification
%-----------------------------------------------------------------------
% Specify how each author name should appear in the author index.  The 
% \paindex{ } should be used to indicate the primary author, and the
% \aindex for all other co-authors.  You MUST use the following
% syntax: 
%
% SYNTAX:  \aindex{LASTNAME, F. M.}
% 
% where F is the first initial and M is the second initial (if
% used).  This guarantees that authors that appear in multiple papers
% will appear only once in the author index.  
%
% EXAMPLE: \paindex{Crabtree, D.}
%          \aindex{Manset, N.}        
%          \aindex{Veillet, C.}        

\paindex{Gianotti, F.}
\aindex{Trifoglio, M.}     % Remove this line if there is only one author

%-----------------------------------------------------------------------
%			Subject Index keywords
%-----------------------------------------------------------------------
% Enter up to 6 keywords describing your paper.  These will NOT be
% printed as part of your paper; however, they will be used to
% generate the subject index for the proceedings.  There is no
% standard list; however, you can consult the indices for past ADASS
% proceedings (http://iraf.noao.edu/ADASS/adass.html). 
%
% EXAMPLE:  \keywords{visualization, astronomy: radio, parallel
%                     computing, AIPS++, Galactic Center}
%
% In this example, the author noticed that "radio astronomy" appeared
% in the ADASS VII Index as "astronomy" being the major keyword and
% "radio" as the minor keyword.


%-----------------------------------------------------------------------
%			       Abstract
%-----------------------------------------------------------------------
% Type abstract in the space below.  Consult the User Guide and Latex
% Information file (http://hea-www.harvard.edu/ADASS/ in the
% Proceedings section) for a list of supported macros (e.g. for
% typesetting special symbols). 

\begin{abstract}          % Leave intact
% Place the text of your abstract here 
The ESA Packet Telemetry and Telecommand Standards accommodate the requirements of a great variety of scientific space missions, thus providing a standard basis for cost-effective and technically compatible developments of on--board and on-ground data handling systems for a wide range of projects. This paper describes the design and the implementation of a detector--independent software, running on Unix platforms, which has been developed for the near real time acquisition, archiving, and basic processing of the ESA based telemetry and telecommand data produced during the on ground testing and calibration of various X-- and $\gamma$--ray space borne detectors. 

\end{abstract}

%-----------------------------------------------------------------------
%			      Main Body
%-----------------------------------------------------------------------
% Place the text for the main body of the paper here.  You should use
% the \section command to label the various sections; use of
% \subsection is optional.  Significant words in section titles should
% be capitalized.  Sections and subsections will be numbered
% automatically. 
%
% EXAMPLE:  \section{Introduction}
%           ...
%           \subsection{Our View of the World}
%           ...
%           \section{A New Approach}
%
% It is recommended that you look at the sample papers, sample1.tex
% and sample2.tex, for examples for formatting references, footnotes,
% figures, equations, html links, lists, and other special features.  

\section{Introduction}

Every scientific payload adopting the ESA Packet Telecommand (TC) and Telemetry (TM) Standards uses the TC and TM source packet structures in order to receive the commands which determine the payload set up and operations, and to send the data generated by its subsystems.
These standards dictate the guidelines for the packet structure, as formed by an header, a data field and a trailer.  They also state the information to be included in the header and in the trailer in order to allow the verification and decoding of the packet content. Among them, the Application Identifier (APID) is required to identify the subsystem which is the destination or the source of the TC or the TM packet, respectively. In addition, for a given APID, the Type/Subtype keywords identify the specific operative mode under which the data have been produced, and, then, the structure of the data contained in the data field. 

Suitable Electrical Ground Support Equipment (EGSE) hardware and software are required in order to test and verify, before launch, all the payload functionality and performances. To this purpose, the EGSE provides the simulation of the relevant missing on--board subsystems, generates and sends to the payload the TC packets, receives and archives all the TM packets. By inspecting in near real time the TM reports and the TM housekeeping packets, the EGSE verifies the correct execution of the TC, and is able to monitor the payload health, as required to support the basic engineering tests.
In addition, a big effort is required to the EGSE software developers in order to allow the EGSE operator to easily verify all the different scientific operative modes, and, in particular, to verify and calibrate the scientific performance of the detectors illuminated with X-- and $\gamma$--ray sources or beams. 
   
\section{DESIGN CONCEPT}

\begin{figure}
\epsscale{0.8}
\plotone{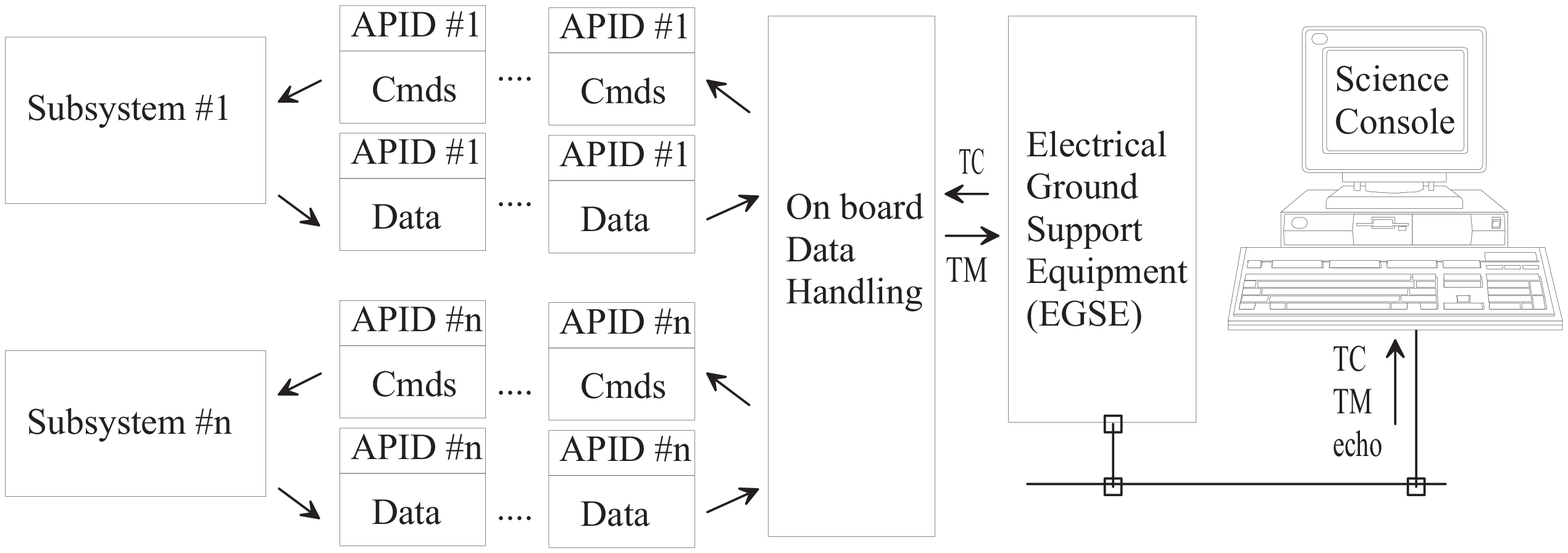}
\caption{TM and TC packets data flow during ground tests.} \label{T1.10-fig-1}
\end{figure}

The use of the ESA TC/TM Standards, allowed us to adopt the common design concept sketched in Figure~\ref{T1.10-fig-1}, where the EGSE is limited to the engineering functionality, while the scientific functionality is provided by the Science Console, which receives in near real time, from the EGSE, a copy of all the TC and TM packets.
The EGSE was procured by the industry, and the Science Console software was in charge of the scientific team responsible for the instrument acceptance test and calibration.

The DISCoS (Detector Independent Science Console Subystem) software we present in the following sections, is the part of the Science Console software which was integrated with various detector specific software written in order to unpack the information contained in the TC/TM packets and to perform the quick look on the scientific data. 
The current version, which is being developed for the AGILE mission (Trifoglio et al. 2000), profits from the experience gained with the previous versions which have been exploited for the XMM--EPIC mission (Trifoglio et al. 1997; Trifoglio et al. 1998), and the INTEGRAL--IBIS mission (La Rosa et al. 1999; Trifoglio et al. 1999).

\section{SOFTWARE ARCHITECTURE AND IMPLEMENTATION}

\begin{figure}
%\epsscale{1.1}
\plotone{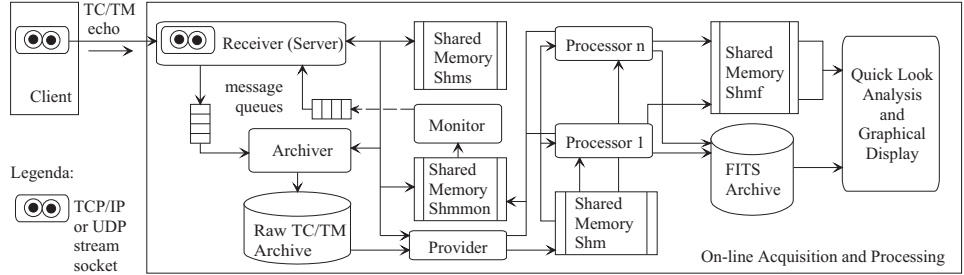}
\caption{The Science Console software architecture.} \label{T1.10-fig-2}
\end{figure}

The DISCoS software consists of the C programs (Monitor, Receiver, Archiver and Provider) which allow the Science Console to acquire, verify and archive in near real time the TC/TM packets in one set of files for each measurement setup, and to reconstruct, either in live or in playback  mode, the various stream of TM scientific packets pertaining to the various operative mode, i.e having the same APID/Type/Subtype.  
Figure~\ref{T1.10-fig-2} sketches how these programs interact together and with the unpacking programs (Processors) and the quick look programs (Quick Look Analysis and Graphical Display) to be written by the DISCoS users.
 
Once started from a shell script, every second the Monitor updates on a screen window the status information and relevant parameters that the concerned programs write on the Shmmon shared memory  by using specific routines. In addition, the Monitor program allows the operator to generates fake start/stop TC, to be used by the Science Console in order to close and open the measurements files independently from the TC packets generated by the EGSE.

The Receiver is the program which interfaces the EGSE through a TCP/IP or UDP socket on an Ethernet LAN 10/100 BaseT. Once started from a shell script, it waits as a Server, and, in the TCP/IP case, establishes with the EGSE a stream socket connection. Hence, using a fork the Receiver generates the Archiver program. 
As long as the connection is active, the Receiver performs a reading loop. On each iteration, by using a non blocking call it reads from the Monitor task message queue the next fake TC packet, if any; and by using a blocking call, it reads from the LAN in order to acquire the next TC/TM packet. 
For each packet, the Receiver inspects the header in order to verify the Application Identifier (APID), the Source Sequence Counter (SSC) and the packet length. 
Hence, by using a blocking call, the Receiver sends through a message queue the packet to the Archiver program, and restarts reading.  
In case a Start/Stop measurement commands is detected, the Receiver waits until the Archiver message queue is empty, then it sends a SIGINT signal to terminate the running Archiver, forks a new Archiver, and eventually restarts reading.

Every time is started, the Archiver opens a new set of raw files:  one to contain all TC/TM packets, and one for the TC and the TM Housekeeping only. Their locations and names are automatically derived from the progressive number (Run ID) which identifies the measurement. An additional suffix is used to identify the files containing the data generated during the instrument configuration. 
By using blocking calls, the Archiver reads each TC/TM packet from the message queue and writes each packet on the local disk using low level  C--Unix Unformatted I/O routines. 
Upon receiving a SIGINT signal from the Receiver, the Archiver completes the reading and archiving of any TC/TM packet from the message queue, closes the raw files, and then terminates itself.

The Provider reads and sorts the TC/TM packets from the raw file. This program is run by a shell script either in live or in playback mode. 
In the former, the raw file name is derived from the current Run ID and the forthcoming  TC/TM packets are read upon receiving the SIGUSR1 signal from the Receiver. 
In the playback mode, the program starts reading from  the raw file which name has been provided by the script. 
For each packet, the Provider verifies the correctness of the packet header, sorts the packet by APID, and writes the packet in the column of the shared memory Shm assigned to the APID. Indeed, this shared memory is managed as a two--dimensional circular buffer capable of containing  some hundreds of packets for each APID.
Information exchanged with the Provider, through the shared memory Shms, allows each Processor to read new TC/TM packets having the required APID as soon as they are available in the shared memory Shm.  A synchronization mechanism guarantees that no TC/TM packet is overwritten until it has been read by the related Processor.
Unless a time out is expired, before overwriting the Provider waits for a SIGUSR2 signal generated by the concerned Processor to notify that the new data have been already read.

In order to allow the interfacing with the detector specific programs, the DISCoS software includes a set of C, Fortran and IDL callable routines. They allow the user to write the Processors as separated programs which have access to the monitor window and to the various source packet streams sorted by APID, without having to deal neither with the EGSE interfacing, nor with the TC/TM packet acquisition, verification and archiving. 
Usually, one Processor program is  run for each APID.
For each sorted stream, the Processor has access to  100\% of the acquired packets, irrespective of both the input rate and the processing to be performed on the packet.
A typical Processor reads the TC/TM packets from the shared memory Shm, verifies and archives the instrument data extracted  from the packet data field in a format suitable for further off--line analysis (e.g. FITS format). Other DISCoS routines allows the Processor to communicate these data, through the shared memory Shmf, to another user's program, written in IDL, which is devoted to quick look purposes. Depending on the selected mechanism, on each call the quick look program receives the unpacked data related to either the next or the last packet processed by the Processor. 
A typical quick look program produces and displays in near real time further data products (e.g. time profile, spectra, and images) allowing the user interaction.

\section{CONCLUSIONS}

The DISCoS software presented herein has demonstrated its effectiveness and re--usability in the Science Consoles developed for several EGSE and Test Equipments which have been designed adopting the ESA Packet Telemetry and Telecommand Standards for the instrument data and commands transport structure. 
The software architecture and implementation has allowed a simple porting from the original HP--UX Unix workstation platform to the Linux PC platforms with the GNU Compiler Collection (GCC).


\begin{references}

% G. La Rosa, G. Fazio, A. Segreto, F. Gianotti, J. Stephen, M. Trifoglio:
% "The EGSE Science Software of the IBIS instrument on-board INTEGRAL satellite"
% AIP Conference Proceedings 'The 5th COMPTON Symposium' (Porthsmouth, NH),
% vol. 510, pp. 693-697, 1999
\reference La Rosa, G. et al. \ 1999, \ Proc. AIP Conference Vol. 510, p. 693-697

% The XMM/EPIC EGSE Science Console Software, ERICE
\reference Trifoglio, M.  et al. \ 1997, \ Proc. of the Fifth Workshop Data Analysis 
in Astronomy, ed.~V.~Di~Gesu' et al., World Scientific, p.233-238

% Ground-calibration GSE for the XMM-EPIC instrument at the Orsay Synchrotron facility 
% SPIE 1998, S.Diego 
\reference Trifoglio, M.  et al. \ 1998, \ Proc. SPIE Vol. 3445, p. 558-565

% Science test equipment for the INTEGRAL PICsIT instrument - SPIE 1999, DENVER 
\reference Trifoglio, M.  et al. \ 1999, \ Proc. SPIE Vol. 3765, p. 572-581

% M.Trifoglio, F.Gianotti, J.B. Stephen, E.Celesti, C.Labanti and A.Traci 
% "The Ground Support Equipment for the Scientific tests and calibration
% of the AGILE instrument" SPIE 2000 - S.Diego
\reference Trifoglio, M.  et al. \ 2000, \ Proc. SPIE Vol. 4140, p. 478-485

\end{references}
\end{document}